\documentclass[sigconf]{acmart}
\usepackage{makecell}
\usepackage{totpages}
\usepackage{xcolor}
\usepackage{multirow}
\usepackage{changepage,threeparttable} 
\usepackage{tabularray}
\usepackage{tabularx, makecell, booktabs}
\usepackage{booktabs}  

\usepackage{graphicx}
\usepackage{array}

\definecolor{ConditionA}{HTML}{CD5C5C}
\definecolor{ConditionB}{HTML}{CD853F}

\AtBeginDocument{%
  }

\begin{document}

\title[HolmeSketcher]{HolmeSketcher: Generative 3D Sketch Mapping for Spatial Reconstruction in Crime Scene Investigation}

\author{Tianyi Xiao}
\authornote{Corresponding Author.}
\email{xiaoti@ethz.ch}
\orcid{0000-0003-1358-3690}
\affiliation{%
  \institution{ETH Zürich}
  \city{Zürich}
  \country{Switzerland}
}

\author{Yizi Chen}
\orcid{0000-0003-1637-0092}
\email{yizi.chen@ethz.ch}
\affiliation{%
  \institution{ETH Zürich}
  \city{Zürich}
  \country{Switzerland}
}
\author{Sidi Wu}
\email{sidiwu@ethz.ch}
\affiliation{%
  \institution{ETH Zürich}
  \city{Zürich}
  \country{Switzerland}
}

\author{Peter Kiefer}
\orcid{0000-0003-4457-0438}
\email{pekiefer@ethz.ch}
\affiliation{%
  \institution{ETH Zürich}
  \city{Zürich}
  \country{Switzerland}
}
\author{Yan Feng}
\orcid{0000-0002-1694-8168}
\email{y.feng@tudelft.nl}
\affiliation{%
  \institution{TU Delft}
  \city{Delft}
  \country{Netherlands}
}

\author{Martin Raubal}
\orcid{0000-0001-5951-6835}
\email{mraubal@ethz.ch}
\affiliation{%
  \institution{ETH Zürich}
  \city{Zürich}
  \country{Switzerland}
}

\renewcommand{\shortauthors}{Xiao, et al.}

\begin{abstract}
Sketch mapping is widely used in crime scene investigation (CSI) to document, interpret, and communicate spatial information. However, it is typically performed on 2D media, which limits its ability to represent 3D spatial relationships. We present \textit{HolmeSketcher}, a generative 3D sketch mapping system that combines a front-end 3D drawing interface with a back-end deep learning pipeline to support object generation and scene reconstruction in extended reality. In a within-subject user study (\(N = 15\)), HolmeSketcher improved the spatial accuracy and interpretability of reconstructed scenes, but with a clear trade-off of higher task load and lower usability compared with paper-based 2D sketch mapping. By integrating findings from the user study and expert interviews (\(N = 3\)), we further derive three design implications for next-generation 3D sketch mapping tools for CSI.
\end{abstract}



\begin{teaserfigure}
\includegraphics[width=\linewidth]{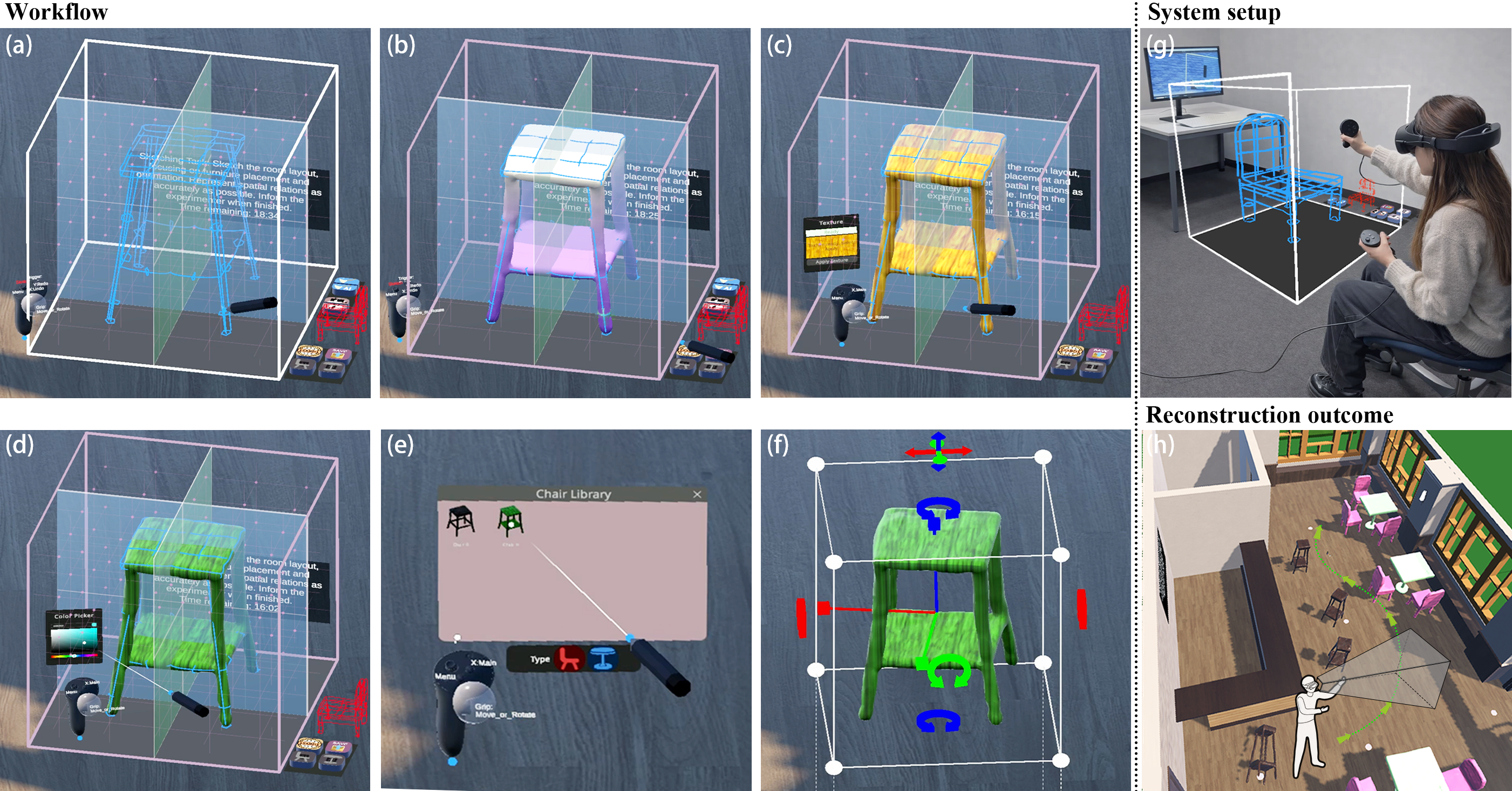}
\caption{\textit{HolmeSketcher} is a generative 3D sketch mapping tool for crime scene reconstruction that supports the externalization of spatial memory during investigative interviews. The workflow includes (a) 3D sketching, (b) an AI-based sketch-to-object pipeline, (c) material generation from voice input, (d) appearance adjustment, (e) object storage and retrieval for scene composition, and (f) object manipulation through scaling, translation, and rotation. Panel (g) shows the system setup. Panel (h) shows the users exploring a reconstructed crime scene composed of the generated chairs and tables.}
\label{fig:teaser}
\end{teaserfigure}

\maketitle

\section{INTRODUCTION}
Crime Scene Investigation (CSI) is a systematic process that involves the identification, documentation, collection, preservation, and analysis of physical evidence from a crime scene to reconstruct events \cite{fisherTechniquesCrimeScene2003, miller2018, ghanem2021}. 
To support CSI, sketch mapping is a widely used externalization practice for scene documentation (Figure~\ref{fig:example sketch maps} (a)), investigative interviews (Figure~\ref{fig:example sketch maps} (b)), suspect identification, and court testimony \cite{brewerPsychologyLawEmpirical2007, dandoModifiedCognitiveInterview2009, bullInvestigativeInterviewing2014, gardner2018, miller2018}. 
The goal of sketch mapping is to create hand-drawn spatial representations \cite{schwering2022}, not only for communication purposes but also to reflect cognition \cite{fanDrawing2023, bainbridgeDrawing2025}.
For these purposes, sketch maps must accurately capture locations, dimensions, and spatial relationships while being produced rapidly to mitigate scene contamination and memory decay \cite{gardner2018, miller2018, xiao2024}.

Traditional sketch mapping mainly relies on paper-based two-dimensional (2D) sketches. However, crime scenes inherently exist in three-dimensional (3D) space, where the vertical dimension plays a crucial role in capturing line-of-sight, ballistics, lighting, and shadows \cite{fisherTechniquesCrimeScene2003, miller2018}. Compressing such information into 2D maps requires advanced skills to accurately project 3D information onto 2D, to depict occlusions and vertical structures, making the process both inefficient and demanding for sketchers \cite{xiao2025sketch2terrain}.
To address this limitation, we present a 3D sketch mapping tool based on Extended Reality (XR), augmented with a generative Artificial Intelligence (GenAI)-driven sketch-to-object pipeline, to support more effective spatial documentation, communication, and reconstruction.

Guided by a formative study with expert interviews and prior works on 3D sketch mapping \cite{xiao2025sketch2terrain, xiao2024, kim2022}, we developed \textit{HolmeSketcher}, an AI-driven generative 3D sketch mapping interface that:
\begin{enumerate}
    \item supports 3D sketch mapping through intuitive externalization and manipulation features;
    \item enables a sketch-conditioned 3D object generation pipeline;
    \item facilitates environmental configuration by adjusting natural and artificial lighting and varying weather conditions; and
    \item supports immersive scene exploration from a first-person perspective via free movement and viewpoint adjustments.
\end{enumerate}

Although the proposed technology is not yet robust enough for deployment in real-world investigations, prior work highlights the growing impact of immersive technologies and scene reconstruction on future CSI workflows \cite{wangVirtual2019, zappala2024}. Motivated by this potential, we conducted a user study to evaluate our interface, where 15 participants first explored a simulated indoor crime scene using virtual reality (VR) glasses and subsequently externalized it from memory through sketching. We addressed the following research questions:

\begin{figure}[ht]
    \centering
    \includegraphics[width=\linewidth]{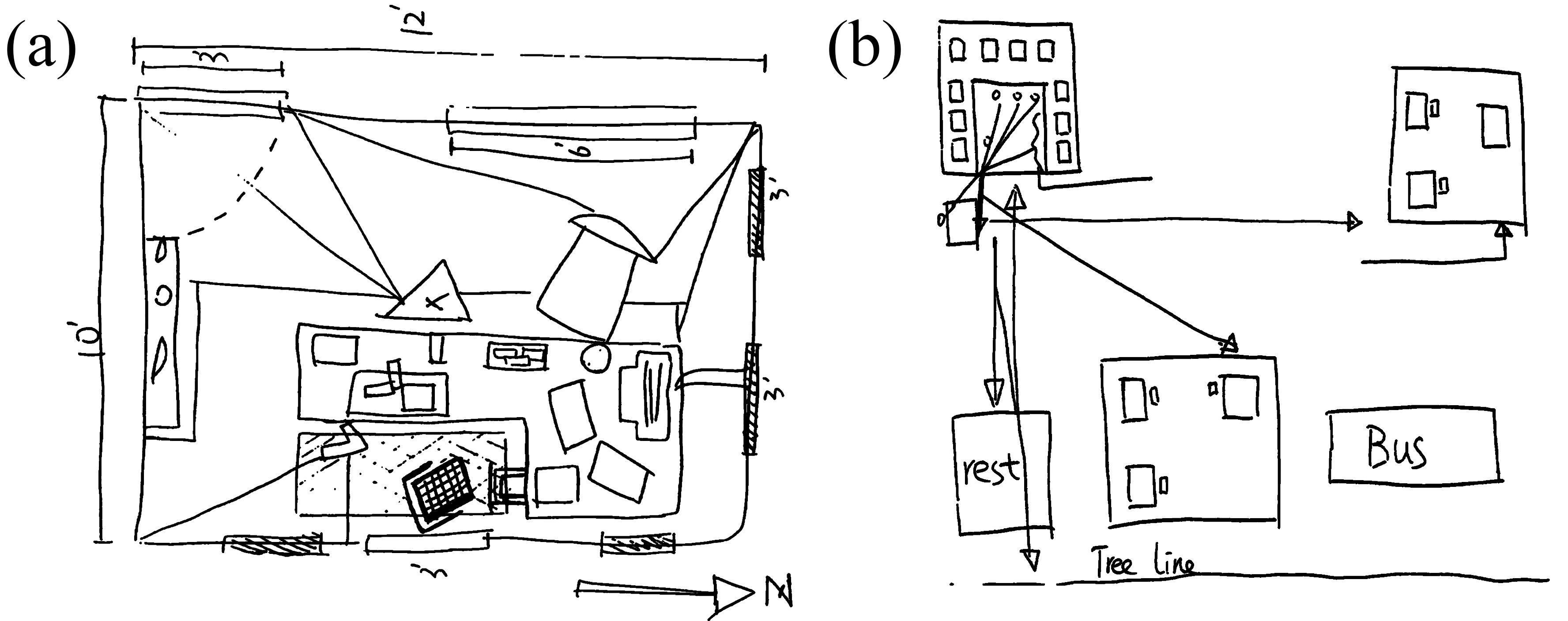}
    \caption{Example sketch maps for (a) scene documentation and (b) witness interviews, redrawn from \cite{luther2023sketching, dutelle2020introduction}.}
    \label{fig:example sketch maps}
\end{figure}


\begin{itemize}
    \item[RQ1.] What are the differences between 2D and generative 3D sketch maps in terms of accuracy and interpretability?
    \item[RQ2.] How does 3D sketch mapping affect usability, user experience, and task load compared to 2D sketch mapping?
\end{itemize}

This study represents a first-of-its-kind step toward understanding the use of 3D sketch mapping in future CSI missions and offers design implications for future tool development.
We contribute:
\begin{enumerate}
    \item The design, implementation, and evaluation of \emph{HolmeSketcher} in an investigative interview use case. 
    \item A set of design objectives for future AI-driven CSI sketch mapping systems, derived from a formative study (N = 9).  
    \item Empirical insights and design implications of future 3D sketch mapping tool for CSI from a user study (N = 15) and expert interviews (N = 3).
\end{enumerate}

\section{BACKGROUND}
\label{Sec:Background}
We first discuss the use of XR and sketch maps in CSI, and we conclude by reviewing prior HCI research on sketch-based modeling.

\subsection{Extended Reality in CSI}
XR spans immersive technologies along the reality–virtuality continuum, including virtual, augmented, and mixed reality (VR, AR, and MR) \cite{milgramTaxonomy1994}. In CSI, these technologies have been increasingly used to support crime scene reconstruction, evidence documentation, training, and courtroom presentation. It is worth noting that few technologies are yet sufficiently robust to survive and support actual CSI use cases. Generally, research is done on prototypes that are aimed at providing the right interface in the future, which is how the present research should be viewed.

In current literature, VR is mainly used to recreate crime scenes for post-event analysis and communication, while AR and MR are more often applied to overlay digital information onto physical spaces, support in-situ inspection, and enable collaborative review of reconstructed scenes \cite{albeedan2023seamless, kim2023digital, cover2017methodology, carew20213d}. Across these applications, XR has been reported to improve spatial understanding, reduce disturbance to physical evidence, and offer new ways for investigators and other stakeholders to examine complex scenes \cite{kim2023digital, mayne2020virtual, pringle2022extended, rinaldi2022virtual}. Recent studies have further explored integrating AI-based techniques, such as object recognition and trajectory analysis, to support scene interpretation and hypothesis development \cite{zappala2024enhancing, rangelov2024impact, amato2020semantic}.

This paper focuses on the use of sketch maps, a long-standing practice in CSI that has received limited methodological attention in previous XR-related research. This research remains significant because spatial understanding and interpretation are cognitively demanding, requiring legal stakeholders to construct and share mental models of the scene for critical reasoning and decision-making. XR systems must therefore support not only accurate reconstruction but also spatial memory, situational awareness, and reasoning. Motivated by this gap, we examine how XR-based crime scene reconstruction relates to spatial perception and memory through sketch mapping, as introduced in the next section.

\subsection{Sketch Maps in CSI}
A sketch map is a hand-drawn spatial representation, typically created on paper, that externalizes an individual’s mental representation of an environment, also referred to as a cognitive map \cite{schwering2022, kim2022, tversky1993}. In CSI, sketch maps function as both cognitive and communication aids across documentation, investigation, and courtroom testimony. They provide a selective representation of scene layout, evidence locations, and spatial relationships that are often difficult to convey through photographs, video, or 3D scans alone \cite{carew2019, gardner2018}. Investigators typically sketch a rough diagram on-site and later refine it into a formal diagram for use in reports and presentations \cite{gardner2018, miller2018}. In court, these sketches serve as shared spatial references for understanding crime contexts and evidence relationships \cite{brewer2004}. Because sketching helps users interpret, organize, and recall spatial information, it is also widely used in investigative interviews, where witnesses or victims draw diagrams to support their accounts and aid memory retrieval \cite{brewerPsychologyLawEmpirical2007, dandoModifiedCognitiveInterview2009, bullInvestigativeInterviewing2014}. 
It can improve recall \cite{dandoModifiedCognitiveInterview2009, dando2020, barlow2011}, elicit richer reports \cite{giolla2017}, enhance statement understanding \cite{luther2023}, and support deception detection \cite{deeb2021, leins2011, vrij2010}.

One major limitation of traditional pen-and-paper sketch mapping is that it produces static 2D representations, which makes it difficult to depict complex 3D structures and dynamic events \cite{xiao2025sketch2terrain, kim2022}. In practice, investigators often compensate by drawing multiple sketches from different viewpoints, such as overview, elevation, and perspective views \cite{ghanem2021}. Others may revisit the scene to add omitted details. However, the former approach is time-consuming, while the latter becomes impractical once the scene has been altered, damaged, or contaminated. Digital mapping tools, such as Microsoft Visio \cite{microsoft2025}, Easy Street Draw \cite{trancite2025}, and Crime Zone \cite{faro2017}, extend these workflows but remain grounded in 2D interfaces, making 3D spatial relationships harder to express and interpret \cite{xiao2025sketch2terrain, wang2019}.

Empowered by immersive 3D sketching in VR/AR, generative 3D sketch mapping offers a promising alternative by allowing users to externalize spatial memory directly in stereoscopic virtual environments and use AI to transform those sketches into multimodal representations, including graphics, text, and 3D shapes \cite{xiao2025sketch2terrain, Xiao2025CoNaviMap}. Building on this concept, we develop an XR tool for spatial reconstruction in the context of investigative interviewing. Our goal is to help diverse users represent and communicate complex 3D crime-scene information more intuitively through sketch-based interaction, introduced in the next section.

\subsection{Sketch-based Modeling in HCI}
\label{subsec:related-3dsketchingmodeling}

Computer-aided sketch-based modeling has been widely studied at the intersection of HCI, computer graphics (CG), and computer vision (CV). In HCI, systems are distinguished by their roles in the design process: 3D sketching tools such as Tilt Brush~\cite{tiltbrush} support artistic expression and early-stage conceptualization, whereas modeling systems such as Gravity Sketch~\cite{gravitysketch} and SimpModeling~\cite{10.1145/3472749.3474791} enable the precise construction of geometry and structure~\cite{10.1145/3126594.3126662}.
Generative approaches further extend these workflows by leveraging algorithms, parametric modeling, and deep learning to explore and optimize design alternatives under given constraints~\cite{ma2021generative, 10.1145/3613904.3642901}.

Early work established freehand sketching interfaces for 3D modeling. Teddy~\cite{igarashi2006teddy} demonstrated how simple 2D sketches can be inflated into 3D forms, while later systems such as CASSIE~\cite{yu2021} improved structural reasoning by constructing curve networks and generating surfaces through constraint optimization. In parallel, research in CG and CV explored \textit{sketch-based 3D object generation}, initially learning deterministic mappings from 2D sketches to 3D shapes~\cite{karpenko2006smoothsketch,wang20223d,chen2023deep3dsketch+,zang2023deep3dsketch+}, and more recently adopting diffusion-based generative models for increased diversity and realism~\cite{bandyopadhyay2024doodle,zheng2023locally}. To address the inherent ambiguity of single-view sketches, prior work introduced viewpoint conditioning~\cite{zhang2021sketch2model,chen2023deep3dsketch,guillard2021sketch2mesh,zheng2023locally} or leveraged multi-view sketches for improved geometric consistency~\cite{lun20173d,delanoy20183d}. Complementing these efforts, recent sketch-to-shape systems such as LAS Diffusion~\cite{10.1145/3592103} and SALAD~\cite{Koo:2023Salad} demonstrate how generative AI can support sketch-driven 3D content creation and completion.

However, most existing systems rely on 2D input, including sketches drawn on tablets or desktop interfaces, as in Teddy~\cite{igarashi2006teddy}, Feather3D~\cite{10.1145/3588427.3595355}, DreamSketch~\cite{10.1145/3126594.3126662}, SimpModeling~\cite{10.1145/3472749.3474791}, and Napkin Sketch~\cite{10.1145/1450579.1450627}. Other approaches use non-sketch inputs, including parametric operations and shape grammars~\cite{stiny1971shape}, as well as, more recently, text and images~\cite{10.1145/3544549.3577043}. In contrast, modern XR head-mounted displays (HMDs) enable users to sketch directly in 3D space, providing a more spatially intuitive interface and reducing ambiguity that inevitably occurs in 2D workflows.
Recent works like~\cite{luo2021data,luo20233d,chen2024rapid,gu2025vrsketch2gaussian} have demonstrated the effectiveness of generative models for 3D shape generation from VR sketches. Based on these advances, we propose the HolmeSketcher, an immersive 3D sketching system that enables users to externalize spatial memory through intuitive mid-air sketching while generating corresponding 3D structures in real time.

\section{\MakeUppercase{Formative Study}}
\label{Sec:Formative Study}
To design our system, we conducted a formative study with three goals: (1) to understand how documentation and interview methods, particularly sketch mapping, affect the criminal justice process; (2) to identify gaps and needs in current practices; and (3) to gather perspectives on VR technologies and 3D sketch mapping. 
\subsection{Method}
We recruited nine experts (E1–E9) with 2 to 20 years of experience from diverse professional backgrounds relevant to crime investigation, including forensic artists, criminal investigators, judges, prosecutors, lawyers, police officers, insurance claims adjusters, and professors, through a screening survey. Semi-structured interviews were conducted online, with questions focusing on current documentation practices, associated challenges, and the potential of XR and 3D sketch mapping for CSI. The full protocol and participant details are provided in Appendix~\ref{appendix:Formative Study Interview Protocol} and Appendix~\ref{appendix:Characteristics of Experts}. Each session lasted one hour, and participants were compensated based on local wage rates. The study protocol received IRB approval from an anonymous university Ethics Commission.

\subsection{Findings}
The first author reviewed the recordings to derive insights. Descriptive coding was used to summarize the findings, which reveal practical limitations and highlight the need for intuitive XR approaches to address them. We summarize them below.

\subsubsection{2D Sketch Mapping Workflow Constrains Complex Scene Documentation}

Crime scene documentation often requires multiple sketches, forcing investigators to translate information across representations. As the complexity of the site increases, this process becomes increasingly time-consuming and labor-intensive. For example, E1 noted that a complex case with indoor and outdoor space transitions may involve ``\textit{more than 15 drawings}'' and take ``\textit{weeks to months}'' to complete. E6 also produced additional sketches when the suspect's and victim's accounts differed, using them to compare accounts and reason about the event. By contrast, 3D interaction can represent spatial structures more directly, reducing the need for multi-angle sketching and mental translation, while offering longer-term benefits despite higher initial effort.

\subsubsection{Static 2D Representation Constrains Scene Understanding}

E6 noted that commonly used documentation methods, such as photographs and sketches, flatten depth, obscure occlusion, and limit spatial understanding to fixed viewpoints. E2 warned that photographs may cause ``\textit{perspective blindness or optical illusions},'' while E1 emphasized the importance of situational awareness, noting that``\textit{you can’t make out the criminal’s face under backlighting}.'' Such contextual information is difficult to capture on static 2D representations. As a result, investigators must imagine the scene from incomplete, viewpoint-limited, and potentially biased records, increasing ambiguity, cognitive load, and the risk of interpretive bias. In contrast, 3D representation enables immersive inspection of spatial layouts from a first-person perspective and across varying eye heights, enhancing spatial perception and promoting empathy.

\subsubsection{Visual-oriented Expression Constrains Scene Interpretation}

E1 described scene interpretation as the process of ``\textit{deduc[ing] events sequence, infer[ring] interactions between physical objects, and reason[ing] about human trajectories}.’’ E9 further noted that physical properties such as material, weight, and mobility are crucial for determining what could have happened. However, current sketch mapping primarily captures visual appearance and cannot represent such physical properties or simulate interactions between objects, requiring investigators to reason through them mentally. As a result, scene interpretation depends heavily on case experience and interdisciplinary knowledge, which are difficult to acquire quickly, as mentioned by E1 and E9. VR could help by making the physical and spatial aspects of scene interpretation more explicit, allowing investigators to simulate material properties, object interactions, event sequences, and even chemical reactions in a reconstructed environment, thereby supporting more intuitive hypothesis testing and reducing reliance on implicit domain knowledge.


\subsection{Elicited Design Objectives}
To address these challenges and guided by the design implications of \cite{xiao2025sketch2terrain, xiao2024}, we generated three high-level design objectives (DOs).
\sloppy
\subsubsection{DO1 — Support users in externalizing complex objects with minimal effort:}
The accurate presentation of physical evidence is crucial to crime scene reconstruction. Generative sketch-to-object pipeline converts rough sketches into precise 3D models. This allows users to focus on memory externalization instead of aesthetic aspects \cite{ma2021generative, 10.1145/3613904.3642901}. AI-based methods adapt to a variety of drawing styles and enable interactive and iterative control \cite{xiao2025sketch2terrain}.

\subsubsection{DO2 — Support compositional scene construction through object manipulation and environmental configuration:}
The spatial arrangement of objects is central to crime scene reconstruction. Object-based composition allows users to incrementally build scenes by placing and manipulating objects, scaffolding the complex task into simpler steps \cite{xiao2024}. Environmental contexts, such as lighting conditions, time-of-day, and weather, are important for hypothesis testing and scene interpretation.

\subsubsection{DO3 — Support embodied scene understanding through immersive exploration:}
Immersive exploration enables users to experience reconstructed scenes from a first-person perspective, allowing them to navigate, revisit, and inspect scenes from multiple viewpoints. This facilitates the interpretation and communication of spatial information, such as layout, distance, and orientation, supporting investigation and court procedures \cite{bolliger2012reconstruction, villa2023virtual}.

\section{\MakeUppercase{HolmeSketcher: System Design} }
\label{Sec:HolmeSketcher Design implementation}

\subsection{Overview}
\begin{figure}[ht]
    \centering
    \includegraphics[width=\linewidth]{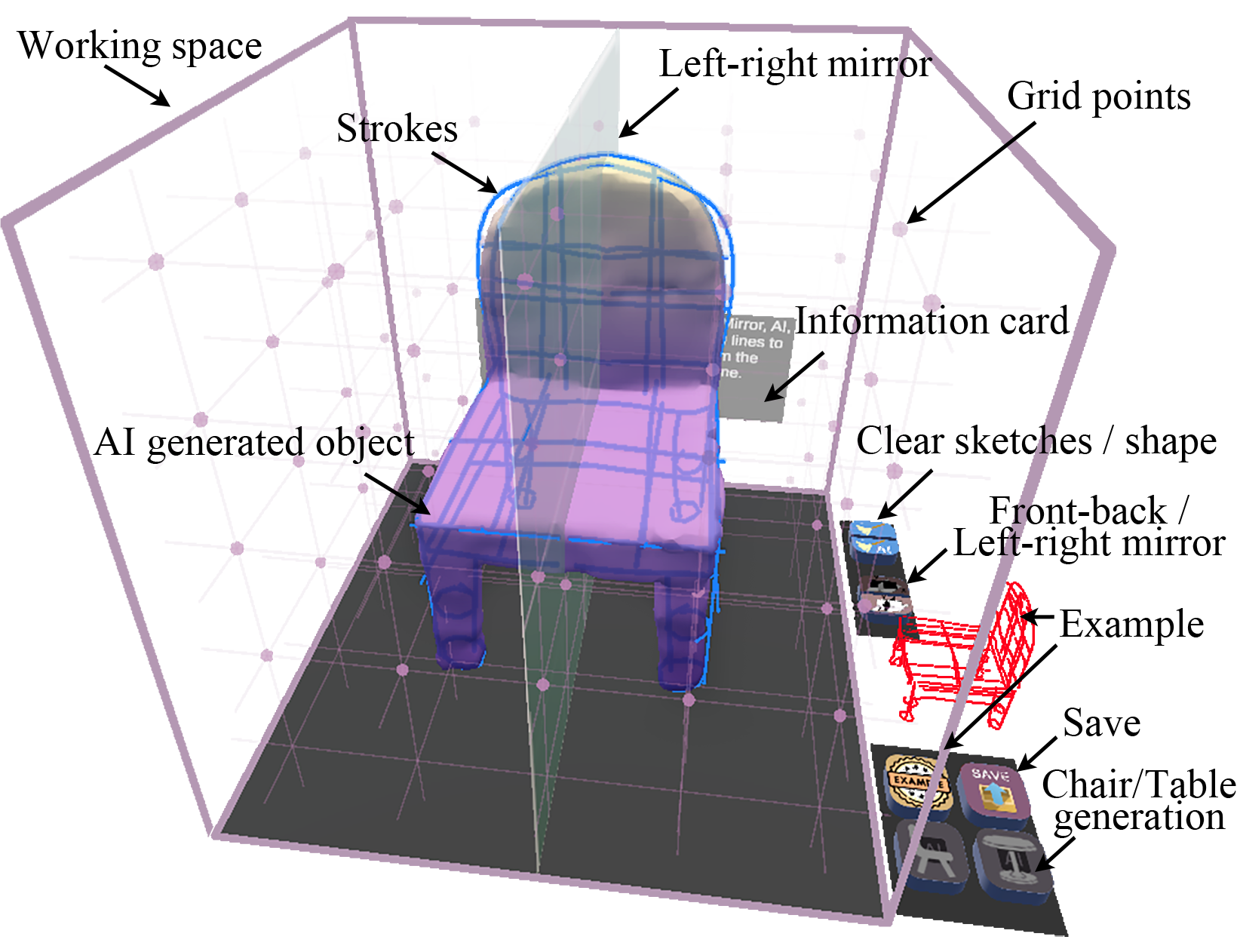}
    \caption{Interface of HolmeSketcher. Users sketch in the workspace to generate 3D objects. Buttons on the right provide supporting functions.}
    \label{fig:Interface design}
\end{figure}

HolmeSketcher is an AI-integrated generative 3D sketch mapping tool that enables diverse users, such as investigators, victims, eyewitnesses, and suspects, to externalize their cognitive maps into immersive virtual environments. Extending traditional sketch maps, it integrates graphics, text, and 3D objects to communicate spatial knowledge more intuitively \cite{xiao2025sketch2terrain}. This work focuses on spatial reconstruction in investigative interviews. The interface design of HolmeSketcher is shown in Figure~\ref{fig:Interface design}. It can support object-level reconstruction and scene-level reconstruction using the sketched objects. Advanced spatial-temporal tasks, such as precise crime event reconstruction and professional guilt inference, are beyond the current scope. Future integration with animation and visual language models (VLMs) could support such tasks by simulating human behavior, object interactions, and spatial reasoning.

\begin{figure*}[tb]
    \centering
    \includegraphics[width=\linewidth]{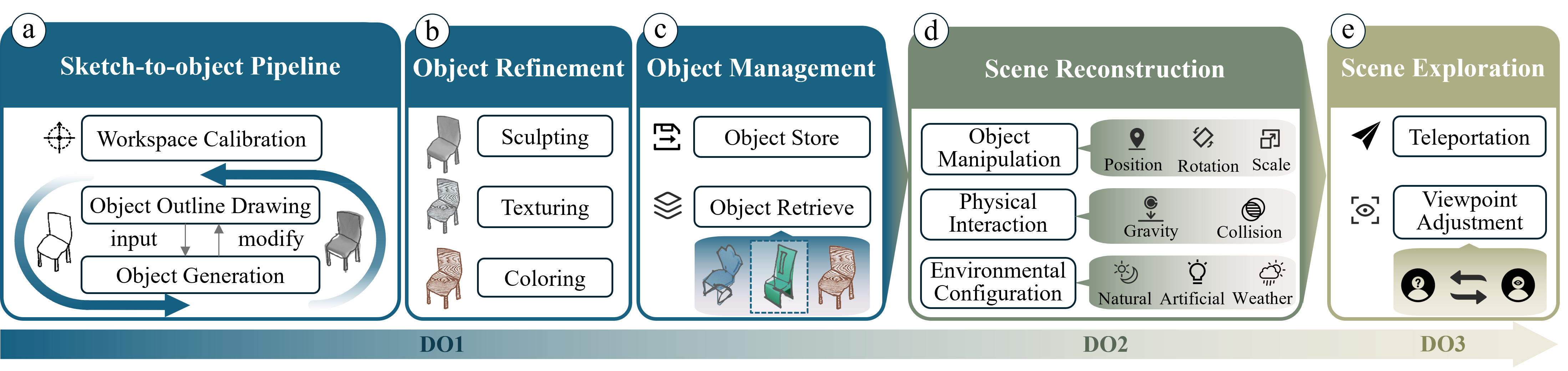}
    \caption{Sketch mapping workflow. (a) Users first align the workspace with the physical environment, then generate objects that match their expectations by iteratively sketching and modifying the 3D outlines. (b) Objects are refined by sculpting surfaces and adjusting appearance through texturing and coloring. (c) Objects can be stored in a library and later retrieved to compose the crime scene. (d) Objects are placed in the scene. Environmental configuration includes lighting and weather. (e) The scene is explored via teleportation, with viewpoint adjustments to match different perspectives (e.g., witnesses or suspects).}
    \label{fig:workflow}
\end{figure*}

\subsection{Workflow \& Features}

The sketch mapping workflow consists of five steps (Figure~\ref{fig:workflow}): (1) sketch-to-object pipeline, (2) object refinement, (3) object management, (4) scene reconstruction, and (5) scene exploration. Steps 1-3 address \textbf{\textit{DO1}} by supporting object-level externalization, generation, and configuration (Figure~\ref{fig:workflow}(a)-(c). Step 4 addresses \textbf{\textit{DO2}} by enabling scene-level reconstruction (Figure~\ref{fig:workflow}(d)). Step 5 addresses \textbf{\textit{DO3}} by supporting scene understanding and exploration (Figure~\ref{fig:workflow}(e)). In the following, we describe each step sequentially.

\noindent\textbf{\textit{Step 1. Sketch-to-object Pipeline.}} Before drawing, users first aligned a 50-cm cubic workspace in front of them by defining a rectangle through its left and right corners \cite{kern20212021, xiao2024}. The workspace includes axis-aligned grid points for spatial reference and depth perception, optional mirror planes for symmetric sketching. Virtual buttons on the right of the workspace provide support functions, an information card shows the current phase and remaining time, and an example sketch to guide the drawing style \cite{yu2021}. 

Users then sketch an object’s main structure using 3D strokes in the workspace (Figure~\ref{fig:teaser}(a)). The system uses sketch sequences as input, encodes them, and sends them to a sketch-conditioned generative model to produce a mesh. The generated mesh is reconstructed in VR, centered in the workspace, and aligned with the sketch  (Figure~\ref{fig:teaser}(b)). Because generation is stochastic, users can choose among multiple outputs or iteratively revise and regenerate.

\noindent\textbf{\textit{Step 2. Object Refinement.}} 
Generated objects can be refined using sculpting, texturing, and coloring tools. Sculpting supports direct mesh editing with brush-based operations such as raise, lower, and smooth. For texturing, users describe desired materials and colors in natural language. An image generator then produces corresponding textures that can be mapped onto the object (Figure~\ref{fig:teaser}(c)). Coloring further enables users to change the color of the texture (Figure~\ref{fig:teaser}(d)).

\noindent\textbf{\textit{Step 3. Object Management.}} The generated objects can be stored in a library (Figure~\ref{fig:teaser}(e)), where users can later retrieve and instantiate them in front of themselves with their original dimensions and appearance. The manipulated objects (in the next step) can also be stored and retrieved in the library.

\noindent\textbf{\textit{Step 4. Scene Reconstruction.}} 
Retrieved objects are placed and arranged within the scene through positioning, scaling, and rotation (Figure~\ref{fig:teaser}(f)). Edited objects can also be stored back in the library for later reuse. We also provide copy and delete functions. The system further supports environmental configuration by enabling adjustments to lighting and weather conditions. Celestial cycle simulation, lighting placement, and weather controls enable accurate reconstruction of natural and artificial illumination, environmental conditions, and the likely time of the crime.

\noindent\textbf{\textit{Step 5. Scene Exploration.}} 
Finally, users can explore the reconstructed scene via teleportation (Figure~\ref{fig:teaser}(h)). Viewpoint adjustment enables embodied scene understanding from multiple perspectives. By adopting the viewpoints of a victim or perpetrator, they can examine differences in the line of sight, distance to objects, and body–object relations, which increase situational awareness and help them infer what actions were physically possible in the scene.



\subsection{Implementation} 
The system consists of a Unity-based VR frontend for 3D drawing \cite{unity} and a backend deep learning pipeline for sketch-conditioned 3D object generation, connected via a Transmission Control Protocol (TCP) client-server architecture. Below, we describe the main components.

\subsubsection{Frontend 3D Drawing System}
The drawing system, derived from CASSIE \cite{yu2021}, provides two modes: mid-air sketching for curved forms and line sketching for straight edges. Strokes are represented as smooth cubic poly-Bézier curves or straight line segments. After sketching, the Ramer-Douglas-Peucker algorithm \cite{saalfeld1999} is applied to smooth and simplify strokes. Users hold the controller like a pen, press the Grip button to start a stroke, and release it to end. Left-right and front-back mirroring duplicate strokes symmetrically across the corresponding plane, supporting efficient and consistent creation of symmetric structures. An axis-aligned point grid is rendered via a Unity shader, with configurable horizontal and vertical spacing \cite{yu2021}.

\begin{figure*}[tb]
\centering
\includegraphics[width=\linewidth]{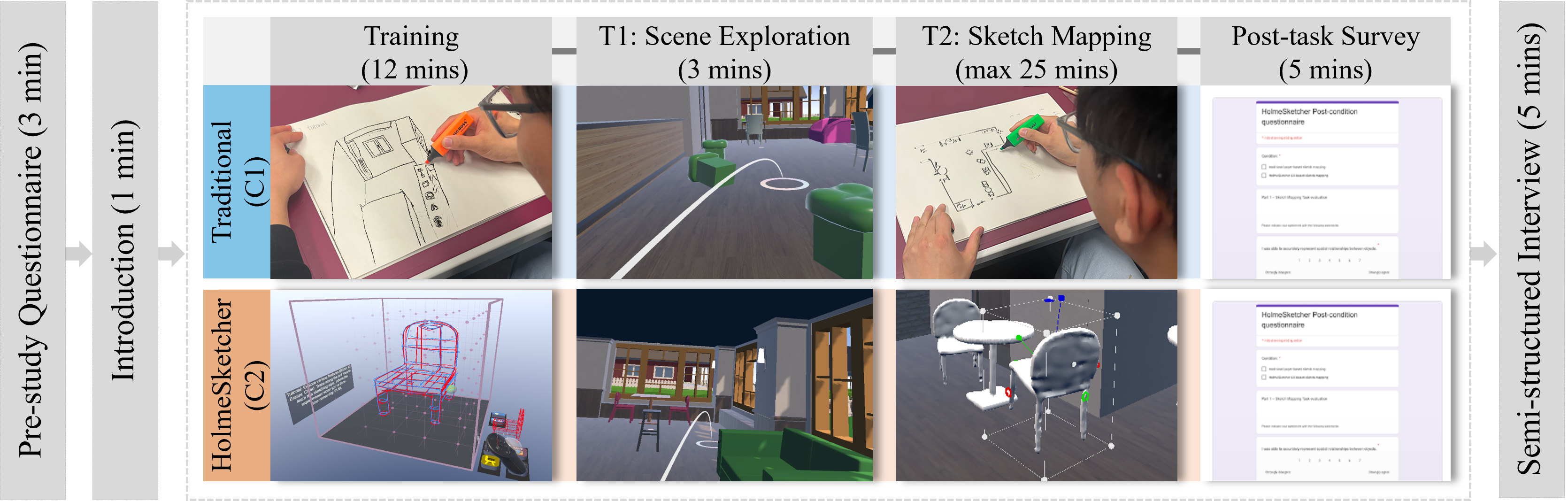}
\caption{The study procedure. Each participant completed both conditions with different stimuli. In C2, we disabled object refinement and environmental configuration to simplify T2 and focus on object sketching and scene reconstruction.
}
\Description{The study comprised five phases: informed consent and a pre-study questionnaire, introduction, training, two tasks followed by a survey, and a post-experiment interview. Each participant completed both conditions with different stimuli.}
\label{fig:Procedure}
\end{figure*}

\subsubsection{Backend Deep Learning Pipeline}
The backend is built upon VRSketch2Shape \cite{chen2025order}, which consists of a pretrained 3D variational autoencoder (VAE) and a diffusion model. The VAE takes signed distance fields (SDFs) as input and projects them into a high-dimensional latent embedding space. The diffusion model is then applied within this latent space, and once the process is complete, the VAE decoder reconstructs the 3D objects. To guide the generation process, the sketch sequences are encoded using a BERT-based architecture with spatial-temporal embeddings, enabling the model to capture both the sequential structure and positional-temporal dynamics of the sketch trajectories. The resulting sketch embeddings are fused with the latent embeddings through a multi-head attention mechanism during the diffusion process. This approach has demonstrated state-of-the-art performance on public datasets in extensive 3D shape evaluations~\cite{chen2025order}.

\subsubsection{Object Refinement and Management}
For post-generation refinement, HolmeSketcher supports mesh sculpting, texturing, coloring, object storage and retrieval, and object manipulation within the immersive workspace. Mesh sculpting is built on the Sculpting Pro Unity plugin \cite{Matej23} and provides three brush-based operations: \textit{Raise}, which expands local surface regions; \textit{Lower}, which depresses them; and \textit{Smooth}, which reduces local irregularities. 

For texturing, users’ verbal descriptions are first transcribed using Whisper \cite{openai2026}. The transcribed text is then processed by the GPT-4.5 language model to extract and formalize texture-related attributes (e.g., metal or transparent), which are used to construct a prompt describing photorealistic textures. This prompt is passed to the GPT Image model \cite{OpenAI2025} to generate the texture, while a color palette is provided to guide and refine the final coloration.

For object management, generated objects and sketches are saved locally using the Easy Save 3 Unity plugin \cite{Moodkie2026}. Sketches are stored as both OBJ files, with strokes rendered as tubular meshes, and JSON files containing stroke coordinates. Generated object meshes are saved as OBJ files together with screenshots captured from a 45$^\circ$ viewing angle for later retrieval from the library page (Figure~\ref{fig:teaser}(e)). The library organizes saved assets by object type. Users can retrieve objects via ray or poke interaction, after which the selected object is spawned in front of them for future placement and resizing.

\subsubsection{Scene Reconstruction and Configuration}
For scene composition through object placement, HolmeSketcher integrates controller-based grabbing and ray-based remote selection using the Meta XR Interaction SDK \cite{metaMetaXRInteraction2026}. 
After an object is selected, a manipulation box with multiple gizmos appears, supporting translation, rotation, uniform scaling, and single-axis scaling (Figure~\ref{fig:teaser}f). Users can reposition objects by directly dragging them or by using a translation handle; rotate them around the x, y, or z axis using axis-aligned handles; uniformly scale them via corner handles while keeping the opposite corner fixed~\cite{Arslan2025}; and scale them along a single axis by dragging an axis-specific box handle. Sliders are additionally provided for remote rotation and resizing. 

To support physically plausible interactions, objects are equipped with Unity's Rigidbody and Collider components, which simulate gravity-induced motion and collision responses between objects.

For scene configuration, we integrate Enviro 3 \cite{hendrikhauptEnviro3Sky} to support reconstruction of scene illumination and weather conditions. Through the menu, users can interactively adjust the time of day with a slider and configure weather states, including clear, cloudy, foggy, rainy, and snowy environments. Users can also place artificial light sources and adjust their distance using the controller thumbstick.

\section{\MakeUppercase{User Study}}
\label{Sec:UserStudy}

This user evaluation investigates the usability, user experience, and task load of HolmeSketcher, as well as the accuracy and interpretability in communicating spatial information in sketch maps, including object dimensions, placement, and relationships, relative to traditional sketching methods. 
The study protocol received IRB approval from the anonymous university Ethics Commission. Participants received compensation at local hourly wage rates.

\subsection{Method}

\noindent\textbf{\textit{Study Design.}}
We employed a within-subjects experimental design to evaluate two conditions: the traditional paper-based 2D interface (C1) and the HolmeSketcher XR-based 3D interface (C2). Each participant completed both conditions on different stimuli in a counterbalanced order based on a Latin square.

\noindent\textbf{\textit{Participants.}}
We recruited 15 participants (10 female, 5 male; age 24–33, \(M=27.07\), \(SD=3.33\)) through researchers' networks, mailing lists, social media, and flyers. Participants had academic backgrounds (6 Bachelor's, 6 Master's, 3 Doctoral degrees) and reported overall moderate prior experience across relevant domains, including painting, gaming, forensic science, 3D navigation, and AR/VR. All participants had normal or corrected-to-normal vision, used their dominant hand. A Mental Rotation Test \cite{vandenberg1978} indicated generally high spatial ability (\(M=4.40/5\), \(SD=1.35\)).

\noindent\textbf{\textit{Apparatus.}}
For C1, we prepared A3 papers, a 12-color set of colored pencils, orange, blue, and green markers, a pen, an eraser, and a ruler. For C2, we used the Meta Quest Pro and its controllers \cite{meta} for interaction. 
A laptop with a 12-core Intel Core i5-12500H processor, 16GB of RAM, and an NVIDIA GeForce RTX 3060 graphics card ran the application. The backend deep learning pipeline is executed on a workstation with a 12-core Intel Core i9-10900X CPU at 3.7 GHz.
All experimental sessions were conducted in an empty lab measuring $3 \times 3$ meters in an anonymous university (Figure~\ref{fig:teaser}(g)).

\begin{figure*}[!t]
\centering
\includegraphics[width=\linewidth]{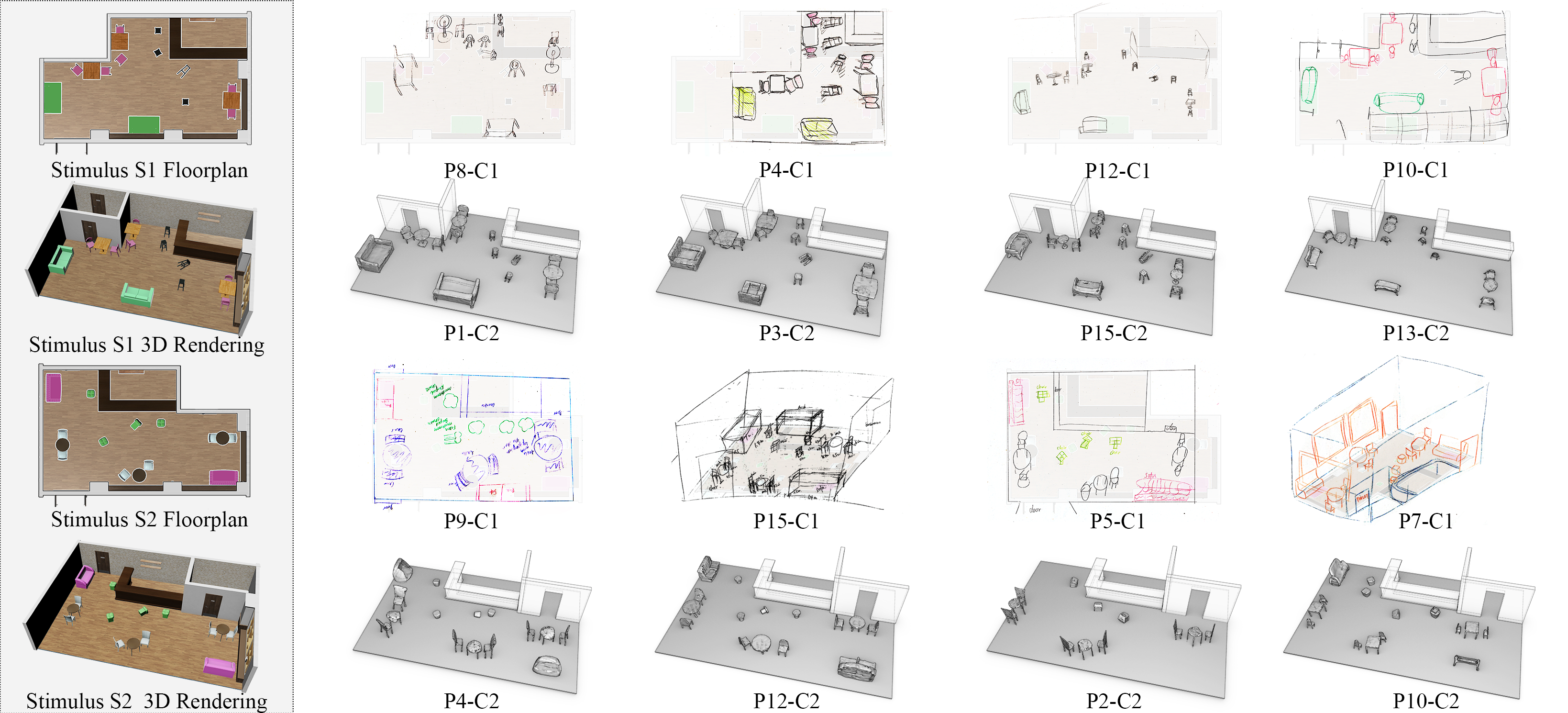}
\caption{The design of the Stimuli, as well as examples of 2D and 3D sketch maps collected in the user study. The 2D sketch maps are overlapping on the corresponding floor plan for visualization purposes.}
\label{fig:floorplan}
\end{figure*}

\noindent\textbf{\textit{Stimuli.}}
\label{Sec:Materials}
We created two similar mock crime scenes as stimuli for the experiment. The base environment was a coffee shop, within which an architect designed and arranged the scene layout. The design was developed through consensus within the research team to ensure that the scene (1) had clearly defined spatial boundaries, (2) contained a variety of object types and sizes, and (3) featured objects with distinct visual characteristics. The number of tables, stools, chairs, and sofas is consistent across both scenes. The stimuli are shown in Figure~\ref{fig:floorplan} (left). The final stimulus design was determined through pilot testing across the two conditions.

\noindent\textbf{\textit{Procedure.}}
\label{Sec:Experiment_procedure} 
The study procedure is illustrated in Figure~\ref{fig:Procedure}. Upon arrival, participants provided informed consent and completed a pre-study questionnaire (Appendix~\ref{appendix:Pre-study Questionnaire}). After a brief introduction, they completed a training session with the assigned interface. In C1, they practiced by sketching their homes on blank paper. In C2, they were introduced to the VR device and completed a tutorial that involved tracing predefined sketches and generating objects.

Following the training, participants first explored a simulated crime scene in VR through teleportation to acquire spatial knowledge (Figure~\ref{fig:Procedure} T1). When the exploration time ended, the furniture disappeared. Task 2 (Figure~\ref{fig:Procedure} T2) required participants to sketch the room layout from memory, including object shape, size, position, orientation, and spatial relationships. This task was not time-constrained. Participants had a maximum of 25 minutes. Figure~\ref{fig:floorplan} shows example sketch maps created in T2. After T2, participants completed a post-task survey (see Appendix~\ref{appendix:post-task-survey}). This procedure was then repeated in the alternate condition with the other stimulus (S1 or S2) to reduce carryover effects. After completing both conditions, participants took part in a post-study interview about their preferences, the perceived strengths and weaknesses of each method, and suggestions for future improvements (Appendix~\ref{appendix:Post-Study Interview Questions}).
\noindent\textbf{\textit{Data Collection and Evaluation Metrics.}}
In C2, sketches were digitally captured in the VR tool, which logged stroke vertices, object mesh, coordinates, and orientations for analysis. In C1, pen-and-paper sketches were scanned and annotated for analysis.

To address RQ1, we evaluated sketch maps in terms of accuracy and interpretability. For interpretability, forensic experts rated the maps using a structured scorecard in terms of spatial layout (SL), object placement (OP), spatial relationships (SR), and object dimensions (OD) (Appendix~\ref{appendix:Interpretability Evaluation Questionnaire}). For accuracy, we measured object position accuracy (OPA), object dimension accuracy (ODA), and object topology accuracy (OTA). To enable comparison, we projected the 3D sketch maps into 2D and applied perspective correction to skewed 2D maps when needed (e.g., Figure~\ref{fig:floorplan} P15-C1 and P7-C1). OPA was computed by optimally matching sketch and ground-truth objects using bipartite matching and then measuring the Euclidean distance between their center points \cite{karp1990optimal}. The distance was converted into a score using \(1/(1+d)\) and min--max normalized to \([0,1]\), with higher values indicating better positional accuracy. ODA was computed using the Intersection over Union (IoU) between sketch and ground-truth object bounding boxes \cite{rezatofighi2019generalized}. To support threshold-based evaluation, we further binarized the IoU values such that objects with an IoU greater than a given threshold were assigned a value of 1, and otherwise were assigned a value of 0. OTA measured whether relative object orientations were preserved along the north–south and east–west axes. The final score was the percentage of correctly preserved orientations, ranging \([0,1]\).

To address RQ2, we administered a post-task survey (Appendix~\ref{appendix:post-task-survey}) that included SUS \cite{brooke1995} for usability, UEQ \cite{laugwitz2008} for user experience, and SIM-TLX \cite{harris2020} for task load, all collected via Google Forms. A semi-structured interview (Appendix~\ref{appendix:Post-Study Interview Questions}) was conducted to elicit preferences. Interviews were audio-recorded, transcribed, and analyzed using open coding followed by thematic analysis.

\subsection{Results}
\label{sec:Result}

\begin{figure*}[!t]
\centering
\includegraphics[width=\linewidth]{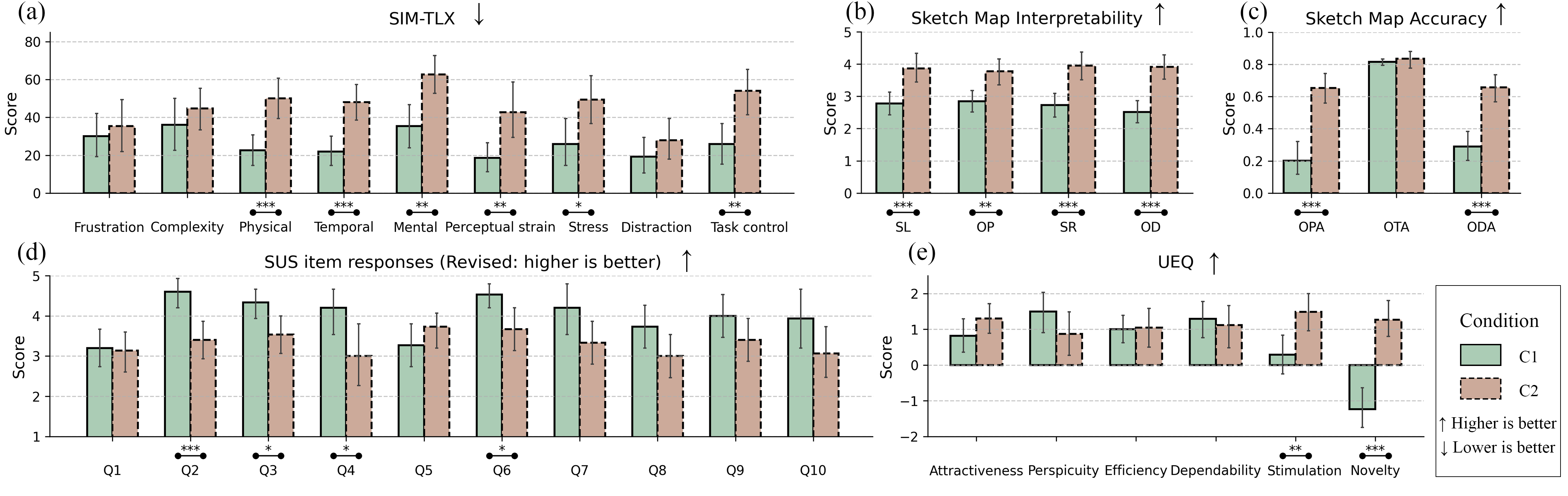}
\caption{Bar plots of user study results. Error bars indicate 95\% confidence intervals. $\ast=p<.05$, $\ast\ast=p<.01$, $\ast\ast\ast=p<.001$.}
\Description{Figure 9: Bar plots illustrating user experience analysis. The error bars represent the 95\% confidence intervals, and statistical significance is indicated by asterisks: * for p<.05, ** for p<.01.}
\label{fig:questionnaire_overview}
\end{figure*}

We conducted 16 sessions, including one pilot session. In the formal study, 15 participants each completed two tasks with two interfaces, yielding 15 valid 2D sketch maps and 15 valid 3D sketch maps. We used t-tests and ANOVA with Tukey’s post-hoc tests for parametric data, and Wilcoxon signed-rank and Mann–Whitney U tests for non-parametric data. Figure~\ref{fig:questionnaire_overview} summarizes the results.

\subsubsection{Sketch Map Analysis.}

Three experts from the expert interviews (Section~\ref{sec:EXPERT INTERVIEWS}) also participated in the interpretability evaluation. C2 scored significantly higher than C1 on all dimensions: SL ($F(1, 88) = 14.06$, $p < .001$), OP ($F(1, 88) = 11.37$, $p < .01$), SR ($F(1, 88) = 16.55$, $p < .001$), and OD ($F(1, 88) = 26.78$, $p < .001$).

For accuracy, C2 scored significantly higher than C1 on OPA ($M_{C1}=0.20$, $M_{C2}=0.65$, $F(1, 28) = 34.02$, $p < .001$) and ODA ($M_{C1}=0.29$, $M_{C2}=0.66$, $F(1, 28) = 31.72$, $p < .001$).

\subsubsection{Post-task Survey.} 
When asked about the fidelity of HolmeSketcher’s generated models, participants reported a mean rating of 4.6 out of 7 (\(SD=1.40\)), showing a moderately positive assessment.

For the SUS, C1 received a significantly higher overall score than C2 ($M_{C1}=75$, $M_{C2}=58$, $F(1, 28)=7.66$, $p<.01$). After reverse-coding the even-numbered SUS items so that higher values reflected better usability, C2 scored significantly lower on Q2, Q3, Q4, and Q6, indicating greater perceived complexity, lower ease of use, more need for technical support, and stronger inconsistency.

For the UEQ, C2 scored higher than C1 on Novelty ($F(1, 29)=40.57$, $p<.001$) and Stimulation ($F(1, 29)=9.07$, $p<.01$).

For the SIM-TLX, C2 was rated significantly higher than C1 on Mental ($F(1, 28)=11.00$, $p<.01$), Physical ($F(1, 28)=15.54$, $p<.001$), and Temporal demand ($F(1, 28)=16.74$, $p<.001$), as well as Perceptual Strain ($F(1, 28)=7.92$, $p<.01$), Stress ($F(1, 28)=6.70$, $p<.05$), and Task Control ($F(1, 28)=11.03$, $p<.01$).

\subsubsection{User Preference}
According to the post-study interview, 60\% of participants preferred C2, while 40\% favored C1. Participants who preferred HolmeSketcher mainly valued its 3D visualization and spatial support, which made it easier to ``recall item sizes and their relative positions'' (P15). Positive feedback described the system as ``super fun and exciting to use'' (P13) and ``straightforward'' (P11). Participants also noted that it enabled a more accurate and detailed representation of the 3D environment and was more accessible to non-experts, as paper-based sketching requires talent and training.'' Some further appreciated features such as symmetrical drawing and object copying. At the same time, they also reported difficulties with precise object placement, including rotation sensitivity and handle selection difficulty. Additionally, the learning curve of 3D drawing makes the system ``unsuitable for a quick sketch.''

Participants who preferred C1 mainly valued their familiarity and simplicity. They described paper-based sketching as straightforward and requiring no additional training or technology, especially when only quick documentation was needed. However, they also noted that ``it can be difficult for people without drawing experience'' and that ``the sketches can look very different among people,'' making the results harder to standardize.

\section{\MakeUppercase{Expert Interviews}}
\label{sec:EXPERT INTERVIEWS}
This expert interview aims to gather professional feedback, assess the potential adoption of HolmeSketcher in real-world CSI practice, and identify opportunities for improvement.
\subsection{Method}
We re-invited experts in the formative study (Section~\ref{Sec:Formative Study}) for semi-structured online interviews. E1, E3, and E9 accepted. Each session included a 5-minute introduction during which the interviewer presented the tool to the participants using images and interaction videos. This was followed by a 55-minute semi-structured interview designed to explore impressions about HolmeSketcher, its potential utility in real-world CSI, comparisons with traditional tools, and suggestions for improvement. A complete list of interview questions is provided in the Appendix~\ref{appendix:Expert Interview}. The study protocol was approved by the IRB board of anonymized university. Participants received compensation at local hourly wage rates. 

\subsection{Findings}
The interviews revealed four insights regarding the application of the 3D sketch mapping tool in CSI practice.

\noindent\textbf{\textit{3D sketch maps improve spatial communication, but do not replace physical reenactment.}} 
All three experts saw clear value in 3D sketch mapping for representing scene layouts, object and character positions, and movement trajectories.  
E1 described the tool as ``\emph{more intuitive than 2D}'' and suggested that it could replace 2D sketch maps in many property-related cases. E3 further noted that 3D representations may be easier to understand for people with lower spatial ability, including elderly users. At the same time, E1 drew a clear boundary around this benefit. For violent crimes involving bodily contact, multiple actors, and rapidly changing actions, virtual reconstruction alone was seen as insufficient. E1 emphasized that such cases still require physical reenactment, because small differences in body characteristics, contact, or force can substantially affect interpretation. In this sense, 3D sketch maps are best understood as a tool for spatial externalization and communication, rather than as a substitute for physical forensic reenactment.

\noindent\textbf{\textit{Local scene completion is more practical than reconstruction from scratch.}} 
Experts noted that full scene reconstruction is uncommon in CSI work. Instead, investigators more often need to complete or restore selected parts of a scene. E9 described this need as ``\emph{local completion of an existing scene},'' rather than creating a scene entirely ``\emph{from scratch}.'' E3 suggested a workflow to first determine the needs of the investigators and prosecutors regarding the level-of-detail and key focus before performing the reconstruction. E1 explained that investigators often compare the current state of a scene with its pre-incident state based on victims' accounts. In burglary cases, for example, this comparison helps reason about offender intent, familiarity with the space, and search patterns. These accounts suggest that reconstruction work in practice is selective and comparative, rather than creating objects from scratch.

\noindent\textbf{\textit{Immersive reconstruction should support narrative reasoning, not only scene depiction.}} 
Experts argued that the value of VR reconstruction extends beyond visual realism. E9 emphasized its role in supporting ``\emph{spatial-temporal reconstruction, storytelling, investigative reasoning, and narrative interpretation}.'' He suggested adding characters, actions, timelines, camera movement, and replay functions. E3 similarly proposed camera-based animation to illustrate how people move through a scene. E1 envisioned layered visualizations of multiple possible escape routes to identify shared locations that may contain critical evidence. Together, these comments point to a broader design opportunity: immersive reconstruction systems should help investigators examine hypotheses and reason through event sequences, not only recreate static objects.

\noindent\textbf{\textit{Real-world adoption is limited by acceptability, learnability, and institutional acceptance.}} 
Although experts recognized the system's practical potential, they identified three barriers to real-world adoption. First, usability for expert users remains a concern. E3 noted the learning demands of VR interaction, limited digital familiarity among senior officers, and the risk of motion sickness. Second, learnability for non-expert users may limit broader use. E9 questioned whether eyewitnesses or other lay users could reliably recall dimensions, communicate spatial relationships, or operate such a tool independently. E9 suggested that expert-assisted use may be more realistic for non-expert users. The third is institutional acceptance. E1 notes that 2D diagrams are widely used in various scenarios within police, prosecutorial, and judicial agencies, are included in standard operating manuals, and are already well-understood by personnel. However, 3D diagrams have not yet been widely adopted and may be met with skepticism. Taken together, these observations suggest that the practical application of 3D sketch mapping in CSI will depend on its ability to integrate into existing workflows and gain acceptance in legal practice.

\section{\MakeUppercase{DISCUSSION}}
\label{sec:discussion}
In this section, we first address our two research questions through a summary of the controlled user study results. We then conclude with limitations and future work. We first summarize our answers to each research question below:

\textbf{RQ1. What are the differences between 2D and generative 3D sketch maps in terms of accuracy and interpretability?}  
Overall, generative 3D sketch maps demonstrate higher accuracy and interpretability than 2D sketch maps. These differences can be understood from both the mapper's and the viewer's perspectives.

From the mapper’s perspective, 2D sketch maps are efficient symbolic representations, but projecting a 3D scene onto a limited 2D canvas makes it harder to preserve object dimensions and spatial layouts accurately, leading to larger distortions in scale and position. For example, in Figure~\ref{fig:floorplan} P9-C1, the tables are drawn disproportionately large relative to the sofa and the overall room. Figure~\ref{fig:floorplan} P4-C1 shows that the 2D sketch map compresses room length, causing global positional distortions. This aligns with the ODA and OPA results, where 3D sketch maps achieve more than twice the ODA and OPA values of 2D sketch maps. In contrast, 3D sketch maps are created in a 1-to-1 scale, enabling users to better reference their position and size relative to surrounding objects. Physics-based interactions further support correct object scaling and placement. Together, these features ensure global spatial correctness.

From the viewer’s perspective, 2D sketch maps require the brain to make additional inferences regarding depth, scale, and height. Overlapping hand-drawn strokes can also reduce the clarity of the sketch. By contrast, generative 3D sketch maps directly encode three-dimensional shapes and preserve shading and occlusion cues, making them easier to recognize and more intuitive to interpret.

\textbf{RQ2.} \textbf{\textit{How does 3D sketch mapping affect usability, user experience, and task load compared to 2D sketch mapping?}} 
Users perceived HolmeSketcher as more creative and engaging, whereas 2D sketch mapping was preferred for its simplicity, familiarity, and ease of use. Participants noted that HolmeSketcher currently requires users to create objects before composing the scene, making the workflow more rigid than 2D sketch mapping, where users can first draft the overall layout and refine individual objects later. Several participants reported that they had already forgotten part of the scene after finishing object creation.

In terms of task load, 3D sketch mapping imposed greater demands because users had to represent the precise object structure, scale, orientation, and position. To reduce the task load, E9 suggested adapting the level of expressive abstraction to different tasks. User feedback suggests that object manipulation is the main source of difficulty and workload. In our setup, the limited physical space prevented natural walking, so users had to rely on remote interaction and teleportation, which were less intuitive than direct grab-based manipulation. Nevertheless, they still appreciate placing objects in a 1-to-1 scale virtual environment, which provides them with a situated experience and enhances their memory.

Overall, 2D and 3D sketch mapping reflect a trade-off between ease of use and map fidelity. 3D sketch mapping provides higher accuracy and interpretability, but at the cost of higher learning demands and task load, making it more suitable for expert users or expert-assisted scenarios. In contrast, 2D sketch mapping remains advantageous for quick, familiar, and low-effort scene representation, especially for non-expert users. These findings indicate that 3D sketch mapping is not intended to replace 2D sketch mapping but to complement it, with each being suited to different use cases.

\subsection{Limitation and Future Work}
\label{Sec:Limitation and future work}

Our current prototype has four main limitations that also point to directions for future work. First, HolmeSketcher currently supports static spatial reconstruction better than dynamic event reconstruction. It is less capable of representing temporal progression, object collision, multi-person interaction, and bodily conflict, a limitation that is particularly salient in violent crime scenarios. Future work should therefore extend the system’s storytelling capabilities with animation support, including characters, actions, timelines, viewpoint replay, and layered visualization to better support narration and comparison across possible scenarios.

Second, the current prototype provides limited support for object placement, which increases the temporal, physical, and mental demands of the sketch mapping task and reduces the overall usability of HolmeSketcher. Future work should improve object placement through scaled-down sandbox interaction (like the World In Miniature feature in \cite{dreyVRSketchInExploringDesign2020, stoakleyVirtual1995}), or larger tracked spaces that better support grab-based manipulation and natural walking.

Third, the current object-level to scene-level workflow may not suit all users' preferences and drawing habits. Future designs should therefore support more flexible workflows. For example, allowing users to first arrange placeholders to establish an overall spatial layout and then create and refine individual objects.

Finally, the prototype remains a lab system rather than a deployable forensic tool. Its interaction cost is still high, especially for users with limited VR experience. The system is therefore better understood as an assistive reasoning tool than as a replacement for physical reenactment or 2D sketch mapping. Future work should reduce the interaction barrier through multimodal input and export formats that better align with institutional and legal requirements.

\section{Design Implications}
We combine findings from the user study and expert feedback to derive three implications for the design of future XR-based 3D sketch mapping systems for CSI.

\noindent\textbf{\textit{Prioritize local scene completion over full-scene reconstruction.}} In practice, investigators are more likely to use sketching to add or replace key objects within an existing space than to reconstruct an entire scene from scratch. They may also need to depict an object's state before and after damage, annotate missing items, and movement paths. Future systems should therefore support rapid global scanning as a reconstruction scaffold, followed by sketch-based local completion, restoration of damaged objects, and visualization of missing items and movement trajectories.

\noindent\textbf{\textit{Prioritize narrative storytelling over static visual depiction.}} 
In CSI practice, investigators need to reason not only about where people and objects are located but also about how actions unfold over time, such as entry and exit sequences, object displacement, and movement paths. Future systems should therefore move beyond static visual reconstruction to support interactive spatiotemporal reconstruction, such as movement trajectories and event timelines. Features such as layered sketching for comparing competing hypotheses, animated trajectory playback, scene walkthrough, and before/after scene comparison may better support investigative reasoning, account verification, deception detection, and hypothesis testing than increased visual realism alone.

\noindent\textbf{\textit{Prioritize adaptive interaction over one-size-fits-all interfaces.}} Future systems should support different levels of interaction complexity and expressive abstraction for different users and scenarios. Investigators and forensic experts may require precise and expressive sketching, editing, and object manipulation, whereas witnesses and victims may benefit more from simplified interaction, abstract sketching, or expert-assisted use. Property crimes may benefit most from scene reconstruction, trace analysis, and movement analysis, whereas violent crimes may require richer representations of body movements, force interactions, collisions, and object damage. To improve real-world usability, future systems can further reduce the learning barrier through multimodal input, sketch-based object retrieval and insertion, and AI-assisted sketch completion.

\section{\MakeUppercase{Conclusion}}
\label{Sec:conclusion}
In this paper, we presented HolmeSketcher, a crime scene reconstruction tool that combines a backend deep learning pipeline with a frontend VR-based sketching interface for sketch-conditioned 3D object generation and scene composition. We conducted a user study with 15 participants and found a clear trade-off between ease of use and map fidelity. Based on the user study and expert interviews, we derived three design implications for future tool design: prioritizing local scene completion, spatiotemporal reconstruction, and adaptive interaction. Although current technologies are still insufficient for real-world deployment, HolmeSketcher is opening new opportunities for both scientific inquiry and practical applications in CSI.

\begin{acks}
This research was funded by the Swiss National Science Foundation (SNSF) under the grant \textbf{3D sketch Maps} (Grant No. 202284).
\end{acks}

\bibliographystyle{ACM-Reference-Format}
\bibliography{references}

\appendix
\section{Formative Study – Interview Protocol}
\label{appendix:Formative Study Interview Protocol}

\subsection*{Part 1 – Current Practices and Challenges}

\noindent\textbf{Background \& Responsibilities}
\begin{enumerate}
    \item Can you briefly describe your responsibilities?
\end{enumerate}

\noindent\textbf{Interviewing \& Investigation Practice (Investigators)}
\begin{enumerate}
    \setcounter{enumi}{1}
    \item Can you describe your typical process for interviewing victims, witnesses, or suspects?
    \item What challenges or limitations do you encounter during investigative interviews?
\end{enumerate}

\noindent\textbf{Documentation \& Representation Practices}
\begin{enumerate}
    \setcounter{enumi}{3}
    \item Can you describe your workflow for documenting a crime scene or reconstructing events?
    \item What types of representations do you use? What are their strengths and limitations?
    \item How do you decide which methods to use for a given case?
    \item What spatial information is most critical to capture?
    \item What challenges or limitations do you encounter during documentation?
\end{enumerate}

\noindent\textbf{Sketching Practice (Forensic Artists)}
\begin{enumerate}
    \setcounter{enumi}{8}
    \item Can you describe your sketching workflow, including types of sketches produced and typical time required?
    \item How do you gather information for sketches (e.g., site visits, photos, measurements)?
    \item What tools do you use, and have you used 3D systems?
    \item What are the strengths and limitations of sketching compared to other documentation methods?
\end{enumerate}

\noindent\textbf{Use of sketch maps in Legal Context}
\begin{enumerate}
    \setcounter{enumi}{12}
    \item How are different forms of evidence (e.g., sketches, photos, 3D models) used in court?
    \item Which forms are most effective or persuasive, and why?
    \item How important is spatial accuracy in legal contexts?
    \item Have you used or accepted digital representations (e.g., VR/3D)? What was your experience?
    \item What challenges arise when presenting or interpreting spatial evidence in court?
\end{enumerate}

\subsection*{Part 2 – Future Vision: VR Technologies}
\begin{enumerate}
    \item What are your initial thoughts on using VR for 3D crime scene sketching?
    \item In what scenarios would a 3D sketching tool be more useful than a traditional 2D method?
    \item What features would you expect from a VR sketch mapping tool for scene reconstruction?
    \item Would you be open to using immersive or interactive media in legal or investigative contexts? What concerns or benefits do you foresee?
    \item Is there anything else we should consider when designing future tools for crime scene documentation and interview?
\end{enumerate}
\section{Characteristics of Experts}
\label{appendix:Characteristics of Experts}
\begin{table}[H]
\centering
\caption{Characteristics of experts (WY = Working Years, G = Gender, Edu = Education).}
\Description{Summary of expert background.}
\label{tab:EXPERT_INTERVIEWS}
\begin{tabular}{p{0.4cm} p{0.4cm} p{2.2cm} p{0.4cm} p{0.8cm} p{0.8cm} p{0.8cm}}
\toprule
\textbf{ID} & \textbf{WY} & \textbf{Role} & \textbf{G} & \textbf{Age} & \textbf{Cases} & \textbf{Edu} \\
\midrule
E1 & 5–9 & \makecell[l]{Forensic artist\\Criminal police} & M & 31–45 & >100 & BSc \\
E2 & 5–9 & Judge & F & 31–45 & >2000 & BSc \\
E3 & $\geq$10 & Prosecutor & F & 31–45 & >2000 & MSc \\
E4 & $\leq$4 & Criminal lawyer & M & 24–30 & <10 & BSc \\
E5 & 5–9 & Criminal lawyer & M & 31–45 & >100 & MSc \\
E6 & $\leq$4 & Traffic police & M & 24–30 & >500 & BSc \\
E7 & $\geq$10 & Insurance adjuster & M & 31–45 & >100 & N/A \\
E8 & 5–9 & Lawyer & M & 24–30 & N/A & BSc \\
E9 & $\geq$10 & Professor & M & 46–60 & N/A & PhD \\
\bottomrule
\end{tabular}
\end{table}

\section{User Study – Pre-Study Questionnaire}
\label{appendix:Pre-study Questionnaire}

\noindent\textbf{Part 1 - Participant Background}

\begin{enumerate}
    \item Age: \_\_\_\_\_\_\_ years old

    \item Gender:  
    [Select one option: Female, Male, Non-binary, Prefer not to say, Other – Write in]

    \item Highest level of education completed:  
    [Select one option: Bachelor's degree, Graduate degree, Other – Write in]
\end{enumerate}

[Instruction:] Rate each statement on a 7-point Likert scale
\textit{(1) Strongly Disagree -- (7) Strongly Agree}

\begin{enumerate}
    \item I have extensive experience in painting.
    \item I have extensive experience in video role-playing games.
    \item I have prior knowledge or interest in forensic science or crime scene investigation.
    \item I feel confident navigating 3D environments.
    \item I have extensive experience using augmented reality (AR) or virtual reality (VR) devices.
\end{enumerate}

\noindent\textbf{Part 2 - Spatial ability Assessment:}  
[User experience was measured using the Mental Rotation Test \cite{vandenberg1978}.]

\section{User Study – Post-Task Survey}
\label{appendix:post-task-survey}


\noindent\textbf{Part 1 – HolmeSketcher Generated Model Fidelity}
[Instruction:] Rate each statement on a 7-point Likert scale
\textit{(1) Strongly Disagree -- (7) Strongly Agree}

\begin{enumerate}
    \item The generated 3D models align well with my expectations.
\end{enumerate}

\noindent\textbf{Part 2 – Usability Assessment:}
[Usability was measured using the SUS questionnaire \cite{brooke1995}.]

\noindent\textbf{Part 3 – User Experience Assessment:}
[User experience was measured using the UEQ \cite{laugwitz2008}, which includes the following dimensions: Attractiveness, Perspicuity, Efficiency, Dependability, Stimulation, and Novelty.]

\noindent\textbf{Part 4 – Task Load Assessment:}
[Task load was measured using the SIM-TLX instrument \cite{harris2020}, which includes the following dimensions: Mental Demands, Physical Demands, Temporal Demands, Frustration, Task Complexity, Situational Stress, Distraction, Perceptual Strain, Task Control, and Presence.]

\section{User Study – Interpretability Scorecard}
\label{appendix:Interpretability Evaluation Questionnaire}

Please evaluate the interpretability of the given crime scene sketch using the 7-point scale below:

\begin{enumerate}
    \item Very Poor – Completely unclear or inaccurate.
    \item Poor – Major details are missing.
    \item Fair – Several ambiguities remain.
    \item Acceptable – Understandable but somewhat incomplete.
    \item Good – Clear with minor issues.
    \item Very Good – Clear and nearly complete.
    \item Excellent – Clear, complete, and easy to interpret.
\end{enumerate}

\noindent Evaluation Aspects:
\begin{enumerate}
    \item[SL] \textbf{Spatial Layout:}  
    The sketch clearly represents the structure and layout of the scene.

    \item[OP] \textbf{Object Placement:}
    The sketch logically and accurately represents the positions of key objects (e.g., table, chair).

    \item[SR] \textbf{Spatial Relationships:}  
    The sketch effectively communicates the positions of the objects relative to one another.

    \item[OD] \textbf{Object Dimensions:}  
    The sizes and proportions of objects appear consistent with expected real-world dimensions.

\end{enumerate}

\section{User Study – Post-Study Interview}
\label{appendix:Post-Study Interview Questions}


\noindent\textbf{Tool Preference}

\begin{enumerate}
    \item Which method would you prefer to use for future crime scene investigation: Traditional or HolmeSketcher?
    \item What factors influenced your preference?
    \item Were there specific contexts where one method was more effective than the other?
    \item What were the main pros and cons of each method?
    \item What improvements would you suggest for HolmeSketcher to enhance its usability or performance?

    \item Is there anything else you'd like to share that could inform the design of future sketch mapping tools for crime scene investigation and documentation?
\end{enumerate}

\section{Expert Interview}
\label{appendix:Expert Interview}
The interviews began with a walkthrough of the tool using videos, images, and a short presentation.

\begin{itemize}
\item[T1] What are your initial impressions of this tool?
\item[T2] Do you see the potential for this technology to support real-world CSI?
\item[T3] Compared to traditional 2D tools (e.g., paper and pen), what are the strengths and limitations of this VR tool?
\item[T4] In what scenarios would this VR tool be more useful than traditional 2D tools?
\item[T5] What features or capabilities would you expect in a fully developed version?
\item[T6] Do you have any concerns about using this VR tool in practice (e.g., usability, training, compatibility)?
\item[T7] What improvements or changes would you recommend?
\item[T8] Is there anything else you would like to share?
\end{itemize}

\end{document}